\documentclass[preprint,showpacs,preprintnumbers,amsmath,amssymb]{revtex4}

\usepackage{graphicx}
\usepackage{dcolumn}
\usepackage{bm}

\begin{document}

\preprint{APS/123-QED}

\title{The spike train statistics for consonant \\
    and dissonant musical accords}

\author{Yuriy V. Ushakov}
\author{Alexander A. Dubkov}%
\affiliation{Radiophysics Department, Nizhny Novgorod State
University, 23 Gagarin Ave., 603950 Nizhny Novgorod, Russia}%

\author{Bernardo Spagnolo}
\affiliation{INFM-CNR, and Dipartimento di Fisica e Tecnologie
Relative, Group of Interdisciplinary Physics, Universita di
Palermo, Viale delle Science, pad.18, I-90128 Palermo, Italy}%

\date{\today}

\newcommand{\erfc}{\mathop{\rm erfc}\nolimits}

\begin{abstract}
        The simple system composed of three neural-like
        noisy elements is considered. Two of them
        ({\it sensory neurons} or {\it sensors}) are
        stimulated by noise and periodic signals with
        different ratio of frequencies, and the third one
        ({\it interneuron}) receives the output
        of these two sensors and noise.
        We propose the analytical approach to analysis of
        Interspike Intervals (ISI) statistics of the
        spike train generated by the interneuron. The ISI distributions
        of the sensory neurons are considered to be known.
        The frequencies of the input sinusoidal signals are in ratios,
        which are usual for music. We show that in the case of small
        integer ratios (musical consonance) the input pair of sinusoids results in
        the ISI distribution appropriate for more regular output spike
        train than in a case of large integer ratios (musical dissonance) of input
        frequencies. These effects are explained from the viewpoint of the proposed
        theory.
\end{abstract}

\maketitle

\section{Introduction}
    Since 1980-th it is well known that noise in physical systems
    doesn't always play a negative role. The phenomena of
    coherence and stochastic resonance are found in different branches of
    science \cite{Gammaitoni}, and the typical field of research regarding the
    constructive role of noise is a wide
    class of neural systems. Indeed, it is very difficult to
    forget about noise, investigating various parts of the central
    or the peripheral nervous system, even if a study is carried
    out in a framework of some mathematical model: noise is inherent for the
    dynamics of the membrane potential (due to ion channels' noise)
    of neural fibers and somas;
    the synaptic junctions exhibit stochastic behavior; each neuron
    receives on the average $10^4$ inputs from its neighbors \cite{Gerstner}
    that in itself requires
    statistical methods of investigation, and etc.
    The motivating question of the given research is: how a signal
    survives in such noisy environment?

    Looking for an approach to the problem we concentrated our attention
    to sensory systems \cite{Longtin, Fishman}. Typically, in
    sensory systems there is a set of neurons (referred to as {\it sensory
    neurons} or {\it sensors}) receiving signals directly from the environment.
    For example, in a simple approximation of the mammals'
    auditory system it is supposed that the basilar membrane
    performs the Fourier transformation on an input sound signal
    \cite{Helmholtz}, and
    the  sensory neurons attached directly to this membrane, percept
    different sinusoidal components (depending on
    coordinates of connection along the membrane)
    of the sound as input. Under driving of
    these signals and the mentioned noise they generate trains
    of short impulses (spikes), which are transmitted to other
    neurons ({\it interneurons}) along neural fibers.

    In a number of studies regarded to the neurodynamics under the noise
    influence the interspike intervals (ISI) statistics is of interest.
    In our model, composed of two sensors (stimulated by sinusoidal
    signals and noise) and one interneuron, we
    consider the ISI distribution (ISID) of each sensor to be
    known (from the previous
    works \cite{Burkitt}) and investigate ISIDs of the output signal
    of the interneuron driven by a mixture of noise
    and the sensors'spike trains
    weighted by coupling coefficients. The system with the similar
    structure has been investigated in Ref. \cite{Balenzuela, Lopera} for
    the purpose of {\it Ghost Stochastic Resonance} (GSR)
    phenomenon detection. The GSR term denotes existence
    of the maximum in the system response at some frequency, which is
    absent in the spectrum of the input signal. The maximum takes
    place at some optimal intensity of noise, which
    affects a system as well \cite{Chialvo}.

    Though we have not investigated this
    phenomenon in the presented work, our topic is closely
    connected with GSR studies due to high complexity (multimodality) of
    interneuron's ISIDs comprising peaks
    inappropriate to input sinusoids' periods or their
    multiples.

    We show how the input signal composed of two sinusoids
    is transformed by the proposed noisy system into different types of spike
    trains, depending on the ratio of input frequencies. Looking for
    the differences in the statistical sense, we find out that the
    output ISIDs for some combinations of frequencies have sharp shapes
    similar to ISIDs of an interneuron driven by a well recognized (on a noise
    background) regular signal. Also, there is another type of the
    output ISID (for the other frequencies combinations), which
    has a blurred shape similar to an ISID of a neuron driven by
    randomly distributed impulses.

    In fact, the difference between "sharp" and "blurred" shapes
    of ISIDs is more quantitative, than qualitative, but this difference
    indicates higher stability to noise of one combination of
    input sinusoids in comparison with another one. Investigation of this
    phenomenon can help to understand which types of input signals
    are able to survive in the noisy environment of the brain, which
    principles control this process, and what it means from the
    perceptional, cognitive, and other points of view.

    On the other hand, in the real life a human deals with relatively simple
    combinations of sinusoidal signals, when listens to music. It
    is well known that musical accords (combinations of tones)
    are classified as {\it consonant} (pleasant, harmonious)
    or {\it dissonant} (unpleasant, disharmonious), depending on the ratio between
    frequencies \cite{Plomp}. Thus, use of musical notations appears to be
    convenient in the context of our work for input signals classification
    purposes. However, we should emphasize that our results
    are obtained using the so-called "just intonation" musical
    accords, i.e. frequencies of input sinusoids are related by
    ratios of whole numbers, that is not appropriate for modern
    music, but is more suitable in the presentation sense.

    It is important also, that the consonance and the dissonance
    of accords are recognized by animals (which never deal with
    music) as well \cite{Fishman}. So, the underlying principles seem to
    be common and fundamental for the auditory neural system of
    mammals. This is the good reason to use the neural-like model
    of the auditory apparatus as the object of research into
    effects related with simple signals (like simple musical accords)
    propagation through a noisy nonlinear environment.

    It should be emphasized, that the "noise benefits" phenomena
    like coherence resonance, stochastic resonance, ghost stochastic resonance, etc. are
    appropriate candidates for a solution of signal propagation
    and signal "survival" problems. But they
    allow to reveal a very particular peculiarities
    of signal propagation through the nonlinear noisy environment
    of neural-like systems and don't provide a full statistical picture.
    So, the main goal of the paper is to present an analytical
    description of principles, which control the statistics transformation
    process for spike trains propagated from one level of neurons
    to another one under the influence of noise.

    In the paper we first describe the chosen model in details.
    After that we propose the analytical description
    applied to the interneuron's ISI statistics. In order to
    prove the theoretical conclusions, we compare them with the
    results of computer simulations. Finally,
    we discuss an agreement of obtained results with the
    hypotheses of the consonance and the
    dissonance in music proposed by Helmholtz (1877) and
    Boomsliter\&Creel (1961).

\section{Model}
    As a basis for the investigated neural-like system we have chosen
    the widely used model called Leaky Integrate-and-Fire (LIF)
    neuron. The input neurons (sensors) are driven by the external
    sinusoidal signals, and the output one (interneuron) receives the weighted
    spikes of the input neurons.

    For simplicity we restrict consideration by a case of two
    sensors. As a result, the set of equations for our system can
    be written down in the following form:
    \begin{equation}
        \left\{
            \begin{array}{l}
                \dot{v}_1 = -\mu_1 v_1 + A_1\cos\Omega_1t+\sqrt{D_1}\xi_1(t),\\
                \dot{v}_2 = -\mu_2 v_2 + A_1\cos\Omega_2t+\sqrt{D_2}\xi_2(t),\\
                \dot{v} = -\mu v + k_1s_1(t) + k_2s_2(t)+\sqrt{D}\xi(t),
            \end{array}
        \right.
        \label{general model}
    \end{equation}
    Here: $v_i(t)$ is the membrane potential of the $i^{th}$
    sensory neuron; $\mu_i$ is the relaxation
    parameter; $A_i$ and $\Omega_i$ are the amplitude and
    the frequency of the corresponding harmonic
    input, respectively; $D_i$ is the sensor's noise intensity;
    $\xi_i(t), (i=1,2)$ are the independent sources of the zero-mean
    $\delta$-correlated
    ($<\xi_i(t)\xi_j(t')> = \delta(t-t')\delta_{ij}$)
    white Gaussian noise (WGN) of the sensors; $v(t)$,
    $\mu$, $D$, and $\xi(t)$ are the membrane potential,
    relaxation parameter, noise intensity, and WGN of the
    output neuron (the third equation (\ref{general model})),
    $<\xi(t)\xi(t')>=\delta(t-t')$, $<\xi(t)>=0$,
    and $<\xi(t)\xi_i(t')> \equiv 0$.

    The LIF neuron doesn't comprise any mechanism of spike
    generation. So, as soon as the membrane potential of any
    neuron of the model
    reaches the threshold value $v_{th}$ we say that the spike is
    generated at the threshold crossing instant of time. The
    corresponding membrane potential is reset simultaneously to
    the initial value: $v_i^0$ for the sensors and $v^0$ for the interneuron.

    The spike trains generated by the sensors and received by the interneuron
    are denoted as
    $s_i(t)=\sum\limits_{j=0}^{N_i(t)}\delta(t-t_{ij}), i=1,2$.
    Each spike train is weighted by the corresponding coupling
    coefficient $k_i$. Spikes are modelled by Dirac $\delta$-functions.
    The instants of time $t_{ij}$ correspond to threshold crossings by the
    sensors' membrane potentials, $N_i(t)$ is the number of spikes
    generated by the $i^{th}$ sensor since the initial time.
    Obviously, the values $t_{ij}$ and their numbers
    $N_i(t)$ are directly related with amplitudes of input
    signals that means the system is nonlinear (by the definition of nonlinearity)
    even though it is not clear from the model Eq.(\ref{general model}).

    All simulation and theoretical results presented in the paper are
    obtained using the following set of constant parameters:
    $\mu_1=\mu_2=1$, $\mu=0.3665$, $D_1=D_2=D=1.6\cdot10^{-3}$, $k_1=k_2=0.98$,
    $v_1^0=v_2^0=0$, $v^0=-1$, and $v_{th}=1$,
    unless stated otherwise.

    For the output neuron the {\it refractory period} ($T_{ref}$) is introduced
    explicitly: this neuron does not respond on any external signal
    after reset until the varying potential $v^0e^{-\mu(t-t_{res})}$ reaches the level
    $v=-0.1$. Hence, the refractory period can be written down in
    the following form:
    $$T_{ref}=\frac{1}{\mu}\ln\left(-10v^0\right).$$
    For the chosen parameters we have $T_{ref}=6.28$.


\section{Theoretical study}
        The first two equations of the system (\ref{general
        model}) are, obviously, independent differential
        equations, modelling the well-known Ornstein-Uhlenbeck
        process with the harmonic inhomogeneity and the reset
        rule. The statistics of interspike intervals in this case
        can be obtained analytically or numerically \cite{Burkitt} and we
        consider it to be known.

        The very important thing is that the spike trains of
        our sensors are non-Poisson ones. These spike trains
        are the input into the third neuron, and it means that the
        dynamics of the output neuron membrane potential is non-Markovian \cite{Cox}.
        Hence, we are compelled to investigate the ISI statistics of the output
        neuron using another analytical approach.

    \subsection{Solutions and presuppositions}
        It is possible to obtain analytical solutions for $v_i(t)$
        and $v(t)$ \cite{Gardiner}:
        \begin{equation}
            \begin{array}{l}
                v_i(t)=\left[v_i(t_{0i})-\frac{A_i}{\sqrt{\Omega_i^2+1}}\cos(\Omega_i
                t_{0i}+\phi_i)\right]\times\\
                \times e^{-\mu_i(t-t_{0i})}+\frac{A_i}{\sqrt{\Omega_i^2+1}}\cos(\Omega_i
                t_{i}+\phi_i)+\sqrt{D_i}\zeta_i(t),\\ \\
                v(t)=v(t_0)e^{-\mu(t-t_0)}+\sum\limits_{i=1}^{2} k_i
                S_i(t)+\sqrt{D}\zeta(t).
            \end{array}
            \label{Membrane solutions}
        \end{equation}
        Here: $S_i(t)=\sum\limits_{j=0}^{N_i(t)}e^{-\mu(t-t_{ij})}$
        is a sum of decaying impulses evoked by spikes of the
        $i^{th}$ sensory neuron; $\zeta_i(t)=\int\limits_{t_0}^t
        e^{-\mu_i(t-t')}\xi_i(t')dt'$ is the colored Gaussian noise
        (Ornstein-Uhlenbeck process) with the variance\newline
        $\sigma^2_i(t)=\left<\zeta_i^2(t)\right>=\frac{1}{2\mu_i}\left(1-e^{-2\mu_i t}\right)$ and
        the probability distribution
        $$w^i_{\zeta}(s)=\frac{1}{\sqrt{2\pi}\sigma_i(t)}\exp\left(-\frac{s^2}{2\sigma^2_i(t)}\right).$$
        For the output neuron we have the same forms of
        $\zeta(t)$, $\sigma^2(t)$, and $w_{\zeta}(s)$. $t_{0i}$ and
        $t_0$ are the reset (spike generation) instants of time
        for the sensors or the interneuron, respectively;
        $\phi_i=\arctan\left(\frac{\Omega_i}{\mu_i}\right)$.

        The temporal realizations of membrane potentials of neurons
        allow us to understand the conditions of spike generation by the output neuron
        and to establish connections between these events and input signals.

        In order to perform the following analysis we utilize three main
        presuppositions:
        \begin{enumerate}
            \item The input harmonic signals are {\it
            subthreshold} to the sensors, i.e. the amplitude $A_i$ and
            the frequency $\Omega_i$ are in such a relation, that
            the signal $A_i \cos(\Omega_i t)$ is not able to evoke
            a spike of the $i^{th}$ sensor without noise ($D_i = 0$). From the
            solutions (\ref{Membrane solutions}) we obtain
            \begin{equation}
                \frac{A_i}{\sqrt{\Omega_i^2+1}} < v_{th}.
                \label{Eq:subthreshold}
            \end{equation}

            \item Only one spike can be generated at
            each period of the harmonic driving force. But, at the same
            time, the spiking on each period is the most probable
            situation, and it means the relatively (to $\Omega_i$) high
            relaxation parameter $\mu_i$. Formally, the condition can be
            written down as: $$\frac{1}{\mu_i} \lesssim \frac{2\pi}{\Omega_i}.$$

            \item Each of coupling coefficients $k_i$ is less than the
            threshold membrane potential value $v_{th}$. It
            means that any separate incoming spike evokes also a
            subthreshold impulse of the output neuron's membrane potential
            $v(t)$, i.e. spike generation is impossible without noise.
            At the same time, the sum of two coefficients is greater
            than $v_{th}$:
            $$
            \left\{
                \begin{array}{l}
                    k_{1,2} < v_{th},\\
                    k_1+k_2 > v_{th}.
                \end{array}
            \right.
            $$
        \end{enumerate}

    \subsection{Probability distribution for the output neuron spike}
\begin{figure}
            \includegraphics[width=7cm]{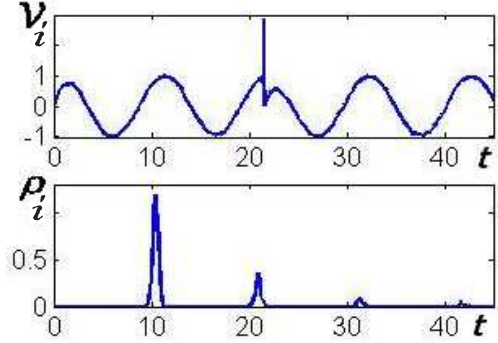}
            \caption{\label{fig:sensor realization} The typical membrane potential realization $v_i(t)$ and the ISI
                distribution of the sensory neuron ($\Omega_i=0.6$,
                $A_i=1.165$). The
                highest probability of a spike after $t=0$ is near
                one period of external force ($t=10$). The
                probability of firing after two, three, etc.
                periods decreases rapidly.}

        \end{figure}

        Let's make some theoretical estimations.
        Initially all three neurons of the system Eq. (\ref{general
        model}) are reset, i.e. $v_1(0)=v_1^0$, $v_2(0)=v_2^0$,
        and $v(0)=v^0$. Since the starting time is $t=0$, we
        measure the first interspike period of the output neuron
        as the first passage time. The first passage time
        probability distributions (FPTPD) are considered to be known
        for the input neurons: $\rho_1(t)$ and $\rho_2(t)$,
        respectively. It means once time is started, we know
        necessary characteristics of a spikes sequence, coming
        from the $1^{st}$ and the $2^{nd}$ neurons to the $3^{rd}$
        (output) one. Spikes of the sensory neurons have a highest probability
        to appear first time at maxima of harmonic driving force
        $(A_i/\sqrt{\Omega_i^2+1})\cos(\Omega_i t_{i}+\phi_i)$.
        They have a narrow probability distribution
        near each of these maxima, and the probability of
        skipping one, two, etc. periods decays exponentially
        (see the Fig. \ref{fig:sensor realization}).

        So, we may analyze a probability $dP_3(t)$ to find the
        $3^{rd}$ neuron spike inside the short time interval $[t, t+dt]$.

        For the chosen parameters there are only 4 probable
        situations in which the output (the $3^{rd}$ one)
        neuron generates a spike:
        \begin{enumerate}
            \item upon receiving a separate spike of the $1^{st}$
            neuron;

            \item upon receiving a separate spike of the $2^{nd}$
            neuron;

            \item upon receiving a $1^{st}$ neuron's spike on a
            background of a $2^{nd}$'s one;

            \item upon receiving a $2^{nd}$ neuron's spike on a
            background of a $1^{st}$'s one.
        \end{enumerate}

        The "separate" spike means that at the time of its
        incoming the $3^{rd}$ neuron's membrane potential $v(t)$
        is driven only by noise, i.e. any previous perturbation over
        the noise level is relaxed.

        The "background" of some incoming spike means that this spike was
        not able to make fire the interneuron, but perturbed
        its membrane potential. This background decays
        exponentially with the decrement $\mu$ until it becomes
        hidden by noise.

        Four described situations exclude each other, so, we may
        take them as independent probabilistic hypotheses. The
        probability of each hypothesis realization is directly
        connected with probabilities of the $1^{st}$ and $2^{nd}$
        neurons' spikes generation on the same time interval
        $[t,t+dt]$, and before it.

        The probability of the $3^{rd}$ neuron spike generation in each of
        these 4 cases depends on the coupling coefficients $k_{1,2}$,
        the noise intensity $D$ and the membrane threshold
        $v_{th}$.

        Hence, we obtain the first term of the contribution
        into the probability of the $3^{rd}$ neuron spike
        generation inside the time interval $[t, t+dt]$ in the
        following form:
        $$dP_1(t)Prob\left\{k_1+\sqrt{D}\zeta(t) \ge v_{th}\right\},$$
        where $dP_1(t)$ is the
        probability of the $1^{st}$ neuron spike generation on the interval $[t, t+dt]$.
        \newline
        $Prob\left\{k_1+\sqrt{D}\zeta(t) \ge v_{th}\right\}$ is the probability of the
        output neuron spike generation under the influence of the $1^{st}$
        neuron's spike (the $1^{st}$ hypothesis probability). For example, if $k_1$ is too small,
        then the $1^{st}$ neuron spike is not able in practice to make fire the
        output neuron.

        The same is applicable to the $2^{nd}$ neuron spikes
        influence, what provides us with the second term.

        For the third hypothesis let's imagine the $2^{nd}$ neuron
        spike comes to the $3^{rd}$ neuron and doesn't make it
        fire. In this case $v(t)$ performs a short "jump" (its
        height is equal to $k_2$) and decays exponentially towards
        zero. According to our presuppositions, during this decay
        we can expect only the $1^{st}$ neuron spike. And it has a
        real chance to make fire the $3^{rd}$ neuron. This
        "real chance" is equal to $Prob\left\{k_1+k_2e^{-\mu(t-t')}+\sqrt{D}\zeta(t) \ge
        v_{th}\right\}$, where $[t',t'+dt]$ is a short interval
        of born of the previous incoming
        spike. It is obvious, that if the instant of time $t'$ is too far
        from the current one $t$, then there is no effect given
        by the previous incoming spike, and in this case a spike at the current
        time $t$ is named the "separate" one. So, we don't need to take into
        account all previous time $t'$. If the previous $2^{nd}$
        neuron's spike doesn't evoke the $3^{rd}$ neuron
        spike, then the first one is totally forgotten by the
        interneuron, when
        $$
            k_2 e^{-\mu(t-t')}=\sqrt{D},
        $$
        i.e. when the noise amplitude becomes equal to the decayed
        impulse (not spike) evoked by the $2^{nd}$ neuron spike.
        By this way we obtain the meaningful period of time to
        integrate over:
        $$
            T_2=\frac{1}{\mu}\ln\left(\frac{k_2}{\sqrt{D}}\right).
        $$

        We also understand, that the whole situation is as seldom
        as high is the probability of the $3^{rd}$ neuron firing under
        influence of a separate spike from the $2^{nd}$ neuron. It
        can be reflected by the factor:
        $$\left(1-Prob\left\{k_2+\sqrt{D}\zeta(t) \ge v_{th}\right\}\right).$$

        This way we obtain the next (third) term of $dP_3(t)$:
        $$
        \begin{array}{l}
        dP_1(t)\left(1-Prob\left\{k_2+\sqrt{D}\zeta(t) \ge
        v_{th}\right\}\right)\times\\
        \times\int\limits_{t-T_2}^{t}
        dP_2(t')Prob\left\{k_1+k_2e^{-\mu(t-t')}+\sqrt{D}\zeta(t) \ge v_{th}\right\}.
        \end{array}
        $$

        The opposite order of 2-spikes sequence (the $4^{th}$ hypothesis) contributes the
        term of the same form with exchanged indexes $2 \leftrightarrow 1$. And the whole expression is:
        \begin{equation}
            \begin{array}{l}
                \rho_3(t)=\rho_1(t)Prob\left\{k_1+\sqrt{D}\zeta(t) \ge
                v_{th}\right\}+$$\\ \\
                +\rho_2(t)Prob\left\{k_2+\sqrt{D}\zeta(t) \ge
                v_{th}\right\}+$$\\ \\
                +\rho_1(t)\left(1-Prob\left\{k_2+\sqrt{D}\zeta(t) \ge
                v_{th}\right\}\right)\times\\
                \times\int\limits_{t-T_2}^t
                \rho_2(t')Prob\left\{k_1+k_2e^{-\mu(t-t')}+\right.\\
                \qquad\qquad \left.+\sqrt{D}\zeta(t) \ge v_{th}\right\}dt'+\\ \\
                +\rho_2(t)\left(1-Prob\left\{k_1+\sqrt{D}\zeta(t) \ge
                v_{th}\right\}\right)\times\\
                \times\int\limits_{t-T_1}^t
                \rho_1(t')Prob\left\{k_1e^{-\mu(t-t')}+k_2+\right.\\
                \qquad\qquad \left. +\sqrt{D}\zeta(t) \ge v_{th}\right\}dt'.
            \end{array}
            \label{p3_1}
        \end{equation}
        Here we switch our attention to the probability densities:
        $\rho_i(t)=dP_i(t)/dt$. And one must
        remember, that everything is valid only for $t > T_{ref}$.

    \subsection{Hypotheses' probabilities}
        In order to make the expression (\ref{p3_1}) more clear, we should
        focus on the coefficients denoted as $Prob\{\dots\}$. The common
        representation of this factor is:
        $$Prob\left\{v(t) \ge v_{th}\right\},$$
        i.e. the probability of the threshold crossing by the
        output neuron membrane potential.

        After expiration of the refractory period and before any
        incoming spike the output neuron membrane potential is
        equal to the Ornstein-Uhlenbeck process realization:
        $v(t)=\sqrt{D}\zeta(t)$.

        Once an external spike is received (e.g. the $1^{st}$ neuron
        spike), $v(t)$ performs an immediate jump to the value
        $k_1+\sqrt{D}\zeta(t)$. Obviously, due to the infinity of the derivation of this
        jump the probability of the $3^{rd}$ neuron spike depends
        only on a current value of the noise realization. That's
        why we may simply write the following:
        $$
        \begin{array}{l}
            Prob\left\{\zeta(t) \ge (v_{th}-k_1)/\sqrt{D}\right\} =
            \int\limits_{(v_{th}-k_1)/\sqrt{D}}^{\infty}
            w_{\zeta}^{st}(s)ds=\\
            \qquad=\frac{1}{2}\erfc\left\{\sqrt{\frac{\mu}{D}}(v_{th}-k_1)\right\}.
        \end{array}
        $$
        Here
        $$w_{\zeta}^{st}(s)=\sqrt{\frac{\mu}{\pi}}\exp\left(-\mu s^2\right)$$
        is the stationary probability distribution of the noise
        amplitude. $\erfc(x)$ is the complementary error function.
        The stationary form is chosen, because the
        refractory period is long enough, and any "jump" of $v(t)$
        without spike generation does not reset the noise component.

        Using the same line of reasoning, it is easy to understand, that in
        a case of 2 incoming spikes close in time we obtain almost
        the same result (e.g. the $1^{st}$ neuron spike comes on the background
        of the decaying  "jump" evoked by the $2^{nd}$ neuron spike):
        $$
        \begin{array}{l}
            Prob\left\{\zeta(t) \ge (v_{th}-k_1-k_2e^{-\mu(t-t')})/\sqrt{D}\right\}
            =\\ \\
            \qquad \qquad=\frac{1}{2}\erfc\left\{\sqrt{\frac{\mu}{D}}(v_{th}-k_1-k_2e^{-\mu(t-t')})\right\}.
        \end{array}
        $$

        When the previous incoming spike and the current one are
        close in time to each other, the difference $(t-t')$ is very
        small. It can be almost equal to zero (simultaneous
        spikes). In such a case we deal with the maximum of $Prob\left\{\zeta(t) \ge
            (v_{th}-k_1-k_2e^{-\mu(t-t')})/\sqrt{D}\right\}$.

        On the other hand, when $(t-t')$ is large (long period
        between the previous and the current incoming spikes), we
        find, that:
        $$
        \begin{array}{l}
            Prob\left\{\zeta(t) \ge
            (v_{th}-k_1-k_2e^{-\mu(t-t')})/\sqrt{D}\right\}\to \\
            \to Prob\left\{\zeta(t) \ge (v_{th}-k_1)/\sqrt{D}\right\},
        \end{array}
        $$
        and this is the minimum of considering probability as a
        function of difference $t-t'$.

\begin{figure*}
            \includegraphics[width=14cm]{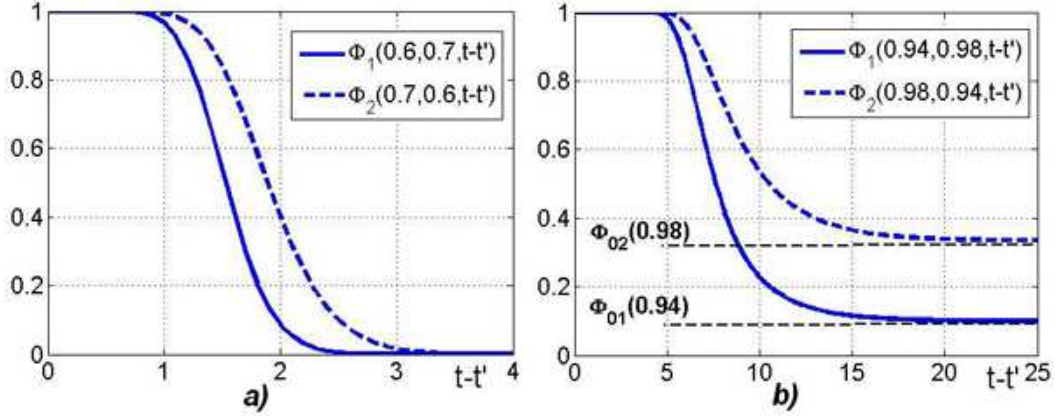}
            \caption{\label{fig:Phi_functions} Interaction functions
            introduced in Eq. (\ref{Eq:PtoPhi}) for weak and strong
            connections: {\bf a)} -- weak connections: $k_1=0.6$,
            $k_2=0.7$. There is a rather sharp boundary of a time
            period, when the second incoming spike is able to
            finish the "work" of the previous one. The separate sensor's
            spike is not practically
            able to fire the output neuron; {\bf b)} -- strong
            connections: $k_1=0.94$, $k_2=0.98$. It is easy to
            see that any separate spike is able to make fire the
            output neuron.
            }
        \end{figure*}

        Denoting
        \begin{equation}
            \begin{array}{l}
                Prob\left\{\zeta(t) \ge
                (v_{th}-k_i-k_je^{-\mu(t-t')})/\sqrt{D}\right\}=\\
                \qquad=\Phi_i(k_i,k_j,t-t')\\
                {\rm and}\\
                Prob\left\{\zeta(t) \ge
                (v_{th}-k_i)/\sqrt{D}\right\}=\Phi_{0i}(k_i)
                \label{Eq:PtoPhi}
            \end{array}
        \end{equation}
        we can rewrite Eq. (\ref{p3_1}) as:
            \begin{equation}
                \begin{array}{l}
                    \rho_3(t) = \rho_1(t)\Phi_{01}(k_1)+\rho_2(t)\Phi_{02}(k_2)+\\ \\
                    \qquad+\rho_1(t)\left(1-\Phi_{02}(k_2)\right)\int\limits_{t-T_2}^t
                    \rho_2(t')\Phi_1(k_1,k_2,t-t')dt'+\\ \\
                    \qquad+\rho_2(t)\left(1-\Phi_{01}(k_1)\right)\int\limits_{t-T_1}^t
                    \rho_1(t')\Phi_2(k_2,k_1,t-t')dt',
                \end{array}
                \label{p3_2}
            \end{equation}
        where
        $$T_{1,2}=\frac{1}{\mu}\ln\left(\frac{k_{1,2}}{\sqrt{D}}\right).$$
        The $\Phi_{1,2}(k_{1,2}, k_{2,1}, t-t')$ factors are depicted at the
        Fig. \ref{fig:Phi_functions} as functions of the time difference $t-t'$.

    \subsection{\label{multiplier} One more important multiplier}
        Regardless of shapes of $\rho_{1,2}(t)$ the FPTPD
        $\rho_3(t)$ must have one important characteristics: if
        the output neuron spike appears at some earlier time,
        then this circumstance decreases the probability of the
        spike in all later moments. We can reflect such a property
        by multiplying $\rho_3(t)$ by
        $$
            \left(1-\int\limits_0^t \rho_3(t')dt'\right).
        $$
        The problem is that our $\rho_3(t)$ may occur to be not
        normalized after the previous calculations (we explain it below, in the section
        \ref{numcheck}). But it is
        possible (without losing of generality and facing with any
        contradictions) first to obtain $\rho_3(t)$ as described above,
        then normalize it, and then multiply the result by the
        mentioned multiplier:
        \begin{equation}
            {\hat \rho}_3(t)={\tilde \rho}_3(t)\left(1-\int\limits_0^t
            {\tilde\rho}_3(t')dt'\right),
            \label{finalRo3}
        \end{equation}
        where ${\tilde \rho}_3(t)$ -- is the normalized
        probability distribution.

    \subsection{A set of the interneuron states}
        \label{states}
        Now we should recall that all previous calculations are
        valid until the first spike is generated by the $3^{rd}$
        neuron. The question is: what happens after?

        At the moment of the $3^{rd}$ neuron's spike generation
        its membrane
        potential $v(t)$ is reset to initial value $v^0$, and the
        interneuron "forgets" all previous history. We suppose
        that the $3^{rd}$ neuron spike is evoked by a
        spike of the $1^{st}$ or the $2^{nd}$ sensor exactly at
        the same moment. Let it be the $1^{st}$
        sensor, which makes fire the
        output neuron. It is also reset to its initial membrane potential
        value $v_1^0$. Consequently, after reset FPTPD $\rho_1(t)$
        has the same shape as it was previously. The other sensor is
        not reset simultaneously with $v_1(t)$ and $v(t)$. Therefore,
        its FPTPD $\rho_2(t)$ is shifted now in comparison with
        the initial situation. The Eq. (\ref{p3_2}) is
        valid, but now it provides us
        with new FPTPD $\rho_3^{(1)}(t)$, where the index $(.)^{(0)}$
        is used for the initial situation. The time is measured
        now since the moment of last spike generation by the
        interneuron. The same is correct after
        each reset of the interneuron:
        one of $\rho_{1,2}(t)$ is similar to its initial form,
        while another one is shifted to the left or right.
        Hereafter we say that the output neuron gets into some {\it
        state} after each reset. These states are defined by
        corresponding shifts of FPTPDs $\rho_{1,2}(t)$ from
        the "viewpoint" of the Eq. (\ref{p3_2}). For detailed
        description see the section \ref{numcheck}. In the case of sinusoidal
        inputs and a finite number of sensors we
        have a finite number of these states. Hence, the
        resulting FPTPD of the output neuron should be
        written as
        $$
            \begin{array}{l}
                \rho_{out}(t)=a_0 {\hat\rho}_3^{(0)}(t)+a_1
                {\hat\rho}_3^{(1)}(t)+\\
                \qquad+a_2 {\hat\rho}_3^{(2)}(t)+\dots+a_{M-1} {\hat\rho}_3^{(M-1)}(t),
            \end{array}
        $$
        where $M$ is a whole number of the interneuron's states. The
        coefficients $a_k$ denote relative frequencies of switching
        into according states. We talk about values of these
        coefficients below in the section \ref{numcheck}.
        Since the distribution ${\hat\rho}_3^{(k)}(t)$ is particular for the $k^{th}$
        state, it would be incorrect to use here $\rho_3^{(k)}(t)$
        (without a hat) and then to perform the same operation as in
        Eq. (\ref{finalRo3}).

        \subsection{Example for different frequencies of sinusoidal inputs}
            \label{exDifFreq}
            Let's take two sensors with input sinusoidal signals of
            different frequencies $\Omega_1 \ne \Omega_2$. We only
            suppose, that these frequencies are in a ratio of some
            integers $m$ and $n$, i.e. $\Omega_1/\Omega_2=m/n$.
            This means that the first $(m-1)$ peaks of $\rho_1(t)$
            don't coincide with the first $(n-1)$ peaks of $\rho_2(t)$.
            And the $m^{th}$ peak of $\rho_1(t)$ coincides with
            the $n^{th}$ peak of $\rho_2(t)$.

            Consequently, the output neuron has
            $M=(m-1)+(n-1)+1=m+n-1$ different possible states against peaks
            of $\rho_{1,2}(t)$, i.e. if it is reset together with
            any spike of the first or the second sensor, then since
            the reset time there is always only one of $M$
            different variants of an incoming spike train
            (the superposition of spike sequences from both sensors)
            with a definite probability density in time for
            each incoming spike.

\section{Numerical experiments }
    \subsection{Consonance and Dissonance in music}
        The sinusoidal signal is considered to be the simplest one
        in an investigation of different systems. May be the first
        level of complication is the linear combination of two
        sinusoidal signals of different frequencies. And here we
        face with the set of very old questions related with
        musical accords.

        The Pythagoreans discovered that the accord of two
        sinusoidal signals sounds pleasant (consonant) if their frequencies
        ratio is $m/n$, where $m$ and $n$ are the small
        integers (e.g. $2/1$, $3/2$, $4/3$) \cite{Plomp}.
        Conversely, if $m$ and $n$ are the large numbers
        (e.g. $45/32$), then the accord sounds dissonant,
        i.e. unpleasant.

        In the context of our investigation it is really
        interesting, how the dissonance and the consonance are
        mapped to ISI distributions.

\begin{figure*}
            \includegraphics[width=12cm]{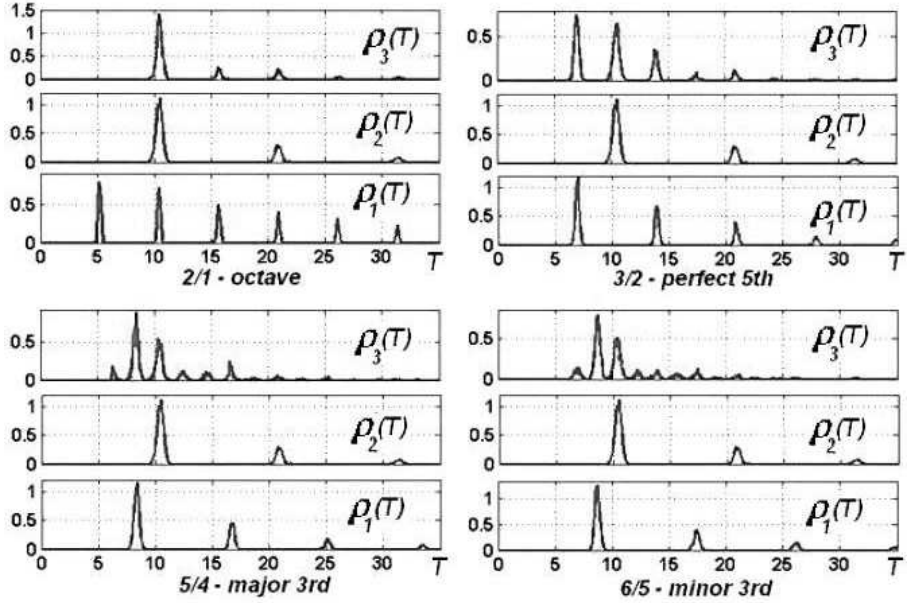}
            \caption{\label{fig:consonants} The consonant accords:
                under each picture there is the
                ratio of frequencies ($m/n$) and the name of the accord used
                in the common musical terminology. All curves are
                obtained through the direct numerical simulation
                of the system Eq. (\ref{general model}) with
                $\Omega_2=0.6$, $A_2=1.165$,
                $\Omega_1=(m/n)\Omega_2$, and $A_1$ according
                the subthreshold input sinusoidal signal condition Eq.
                (\ref{Eq:subthreshold}).}
        \end{figure*}

\begin{figure*}
            \includegraphics[width=12cm]{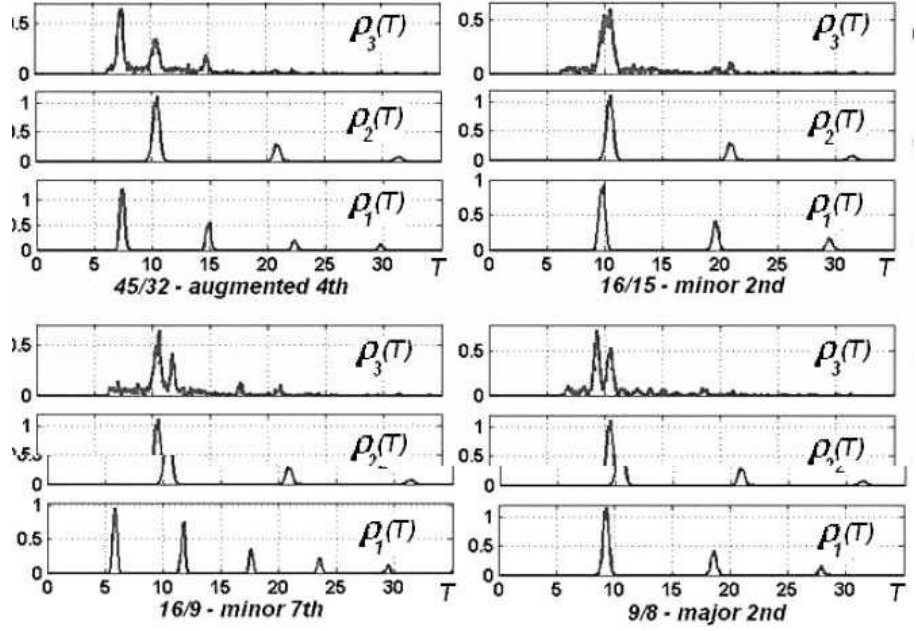}
            \caption{\label{fig:dissonants} The dissonant accords:
                under each picture there is the
                ratio of frequencies ($m/n$) and the name of the accord used
                in the common musical terminology. All curves are
                obtained through the direct numerical simulation
                of the system Eq. (\ref{general model}) with
                $\Omega_2=0.6$, $A_2=1.165$,
                $\Omega_1=(m/n)\Omega_2$, and $A_1$ specified by
                the subthreshold input sinusoidal signal condition Eq.
                (\ref{Eq:subthreshold}).}
        \end{figure*}

        In the Figs. \ref{fig:consonants} and \ref{fig:dissonants}
        there are the distributions for
        consonant and dissonant accords, respectively.
        It is easy to see the higher integers $m, n$ the
        regularity less in an appropriate distribution of ISI,
        although the structure of the input signal is in principal
        the same: two sinusoids. These curves are obtained
        through the direct numerical
        simulations of the system Eq. (\ref{general model}). The
        theoretical part of our work is focused on building a basis
        for the simulations results.

    \subsection{Verification of theoretical conclusions }
    \label{numcheck}
        The formula (\ref{p3_2}) is obtained under a set of
        assumptions. So, this theoretical result should be compared
        with the results of numerical experiments. Here we present the
        idea of usage of the expression and check its validity.

        Let's take, for instance, the "Perfect $4^{th}$" accord, which consists
        of 2 sinusoids of frequencies related by the ratio
        $\Omega_1/\Omega_2=4/3$.

\begin{figure}
            \includegraphics[width=8.8cm]{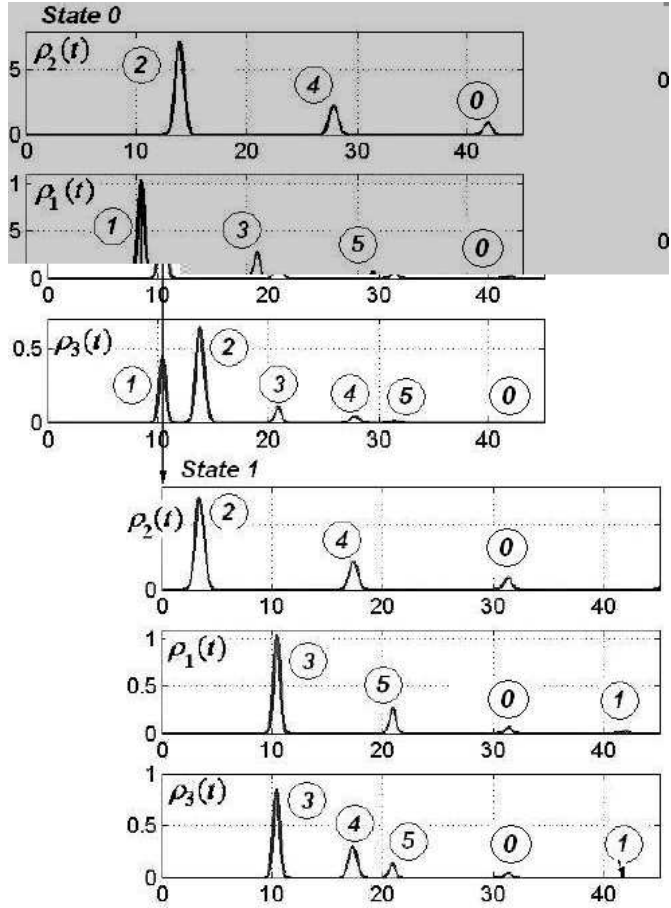}
            \caption{\label{fig:43proc} FPTPDs of three neurons in the states 0 and
                1. This is the example of the probable transition between
                two states for the case of input sinusoids with frequencies
                related by ratio $4/3$ (the Perfect $4^{th}$ accord).
                $\rho_2(t)$ is shifted in the State 1 in comparison with $\rho_2(t)$
                in the State 0. As a result, $\rho_3^{(0)}(t)$ and
                $\rho_3^{(1)}(t)$ are different. All possible
                states of the interneuron are: State 0, State 1,
                \ldots, and State 5. The peaks of $\rho_{1,2}(t)$ are
                marked by numbers in circles in order to establish
                the correspondence between them and the
                interneuron states.
                }
        \end{figure}

        The FPTPDs $\rho_1(t)$ and $\rho_2(t)$ are known to us
        from the numerical simulations of sensors, for example.
        If the figures of these distributions are placed in
        a column (Fig. \ref{fig:43proc}, State 0), then it is easy to
        see, that the $3^{rd}$ peak of $\rho_2(t)$ coincides with
        the $4^{th}$ peak of $\rho_1(t)$. All other peaks
        don't coincide, and (as it
        is explained in the section \ref{exDifFreq}) here we have $4+3-1=6$
        different possible states of the $3^{rd}$ neuron: State 0, State 1,
        \ldots, and State 5.
        Let's establish the correspondence between these states
        and the peaks of $\rho_{1,2}(t)$ as it is shown on the Fig.
        \ref{fig:43proc}: numbers of the states are placed into circles.
        Area under each peak means the probability to
        find an incoming spike at the defined short period of time. If
        this spike evokes the spike of the $3^{rd}$ neuron,
        it is switched into the appropriate state.

        Initially all three neurons are reset. This is
        the $0^{th}$ state. The most probable and close in time spike comes from the
        $1^{st}$ sensor (Fig. \ref{fig:43proc}). If this spike makes fire the
        interneuron, then it is switched into the $1^{st}$
        state, where the most probable and close spike comes from
        the $2^{nd}$ sensor. Obviously, this spike (if it is
        generated) comes during the refractory period, so, the
        closest valuable spike in the $1^{st}$ state comes again
        from the $1^{st}$ sensor and has the possibility to switch
        the interneuron into the $3^{rd}$ state, and etc.

        Here we recall the section \ref{multiplier} and notice
        that the peak of $\rho_1(t)$ or $\rho_2(t)$, which appears
        during the refractory period, is the main reason why we
        may find $\rho_3(t)$ to be not normalized in Eq.
        (\ref{p3_1}). This "invisible" to the output neuron peak
        does not contribute into the $\rho_3(t)$ peaks, but it
        is the big meaningful part of an appropriate normalized sensor's
        FPTPD.

        The analysis of $\rho_3(t)$ peaks in each state shows
        that in the case of musical accords and strong connections
        the interneuron gets into all possible $m+n-1$ states almost uniformly, i.e.
        all states make almost equal contributions into the common
        $\rho_{out}(t)$.

        Therefore, the simplest way to obtain the final output distribution is
        to directly sum all ${\hat\rho}_3^{(k)}(t)$ and then to
        normalize this result. In other words, all coefficients $a_k$
        (see the section \ref{states}) can be set to unit:
        \begin{equation}
            \rho_{out}(t)=\frac{\sum\limits_{k=1}^M {\hat\rho}_3^{(k)}(t)}
            {\int\limits_0^{\infty}dt'\sum\limits_{k=1}^M {\hat\rho}_3^{(k)}(t')}.
            \label{RhoOut}
        \end{equation}

\begin{figure}
            \includegraphics[width=8.8cm]{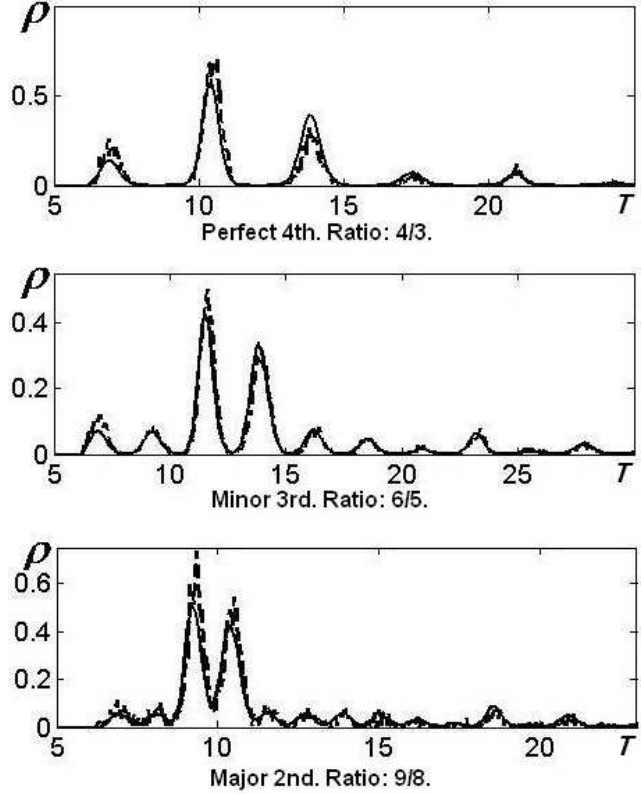}
            \caption{\label{fig:results} ISI distributions of the output neuron for
            different accords. Solid line is the theoretical result.
            Dashed line is the distribution obtained throughout the
            direct numerical simulation of the system (\ref{general model}).
            The parameters are $\mu_1=\mu_2=1$, $\mu=0.3665$, $k_1=k_2=0.97$,
            $D_1=D_2=D=1.6\cdot10^{-3}$. {\bf Perfect 4th:} $A_1=1.165$,
            $\Omega_1=0.6$, $A_2=1.085$,
            $\Omega_2=0.45$.
            {\bf Minor 3rd:} $A_1=1.125$, $\Omega_1=0.54$,
            $A_2=1.085$, $\Omega_2=0.45$.
            {\bf Major 2nd:} $A_1=1.2$, $\Omega_1=0.675$,
            $A_2=1.165$, $\Omega_2=0.6$.}
        \end{figure}

        These approximate conclusions provide $\rho_{out}(t)$
        curves very similar to ones obtained through direct
        numerical calculations of the system (\ref{general
        model}). The examples of compared results are shown in the
        Fig. \ref{fig:results}.

\section{The calculation algorithm}
    Summarizing the previous sections, let's present the described
    theoretical approach in the form of the calculation algorithm.
    Thus, in order to obtain the interneuron's ISID curve under
    chosen parameters of the system Eq. (\ref{general model})
    we should perform the following steps.
    \begin{enumerate}
        \item To obtain the sensors' FPTPDs $\rho_{1,2}(t)$ using
        the direct numerical simulation of the system Eq.
        (\ref{general model}) without the
        interneuron, or theoretical approaches described in Ref.
        \cite{Burkitt}.

        \item To find all possible {\it States} of the interneuron against
        the peaks of $\rho_{1,2}(t)$ (see Fig. \ref{fig:43proc} for
        example).

        \item To calculate $\hat\rho_3^{(i)}(t)$ in each State
        using Eqs. (\ref{p3_2}, \ref{finalRo3}).

        \item And, finally, to sum and normalize the calculated
        $\hat\rho_3^{(i)}(t)$ in accordance with Eq.
        (\ref{RhoOut}).
    \end{enumerate}

    Despite of the relative complexity of the algorithm, its usage decreases
    consumption of resources necessary for smooth interneuron's
    ISID obtaining. It allows also to perform fast estimations and
    provides anyhow the consistent theoretical description of the
    noisy nonlinear system Eq. (\ref{general model}).

\section{Hypotheses of consonance and dissonance}
    There is a few of main hypotheses explaining why animals,
    including humans, feel harmony or disharmony listening to
    different tones combinations. We suppose that the input
    signals, which are transformed into spike trains with blurred,
    i.e. noise-like Interspike Intervals Distributions, are felt
    unpleasant (dissonant, inharmonious) due to the analysis,
    recognition and survival in noisy environment of the brain
    problems.

    Let's try to understand the correlation between that viewpoint
    and some other hypotheses of the dissonance.

    Helmholtz (1877) \cite{Helmholtz} proposed the notion that
    dissonance arises due to beating between adjacent harmonics
    of complex tones. In effect, dissonance arises due to rapid
    amplitude fluctuations.

    It is possible to prove mathematically that, if the input frequencies into our
    system are in ratio $\Omega_1/\Omega_2=m/n$ (where $m > n$),
    then the minimal distance between peaks of $\rho_3^{(i)}(t)$ is
    $T_{min}=(2\pi/\Omega_2)/m=T_2/m$ (see
    the Fig. \ref{fig:43proc}: the peaks 1, 2 of the $\rho_3^{(0)}(t)$,
    and the peaks 4, 5 of the $\rho_3^{(1)}(t)$) that defines
    the distance between peaks of the final ISID $\rho_{out}(t)$.
    That's why the sufficiently high value of $m$ means the blurred
    ISID, typical for dissonant accords. In such a way, we show that
    even for pure input tones, and even if
    they are not close in frequency in order to produce beats,
    we may feel the dissonance. Hence, the hypothesis by
    Helmholtz continues to be correct, if
    we look at the minimal distance among all peaks of $\rho_1(t)$
    and $\rho_2(t)$, and not only the firsts ones, which show the
    difference between the input tones' frequencies.

    Another theory is the Long Pattern Hypothesis from Boomsliter
    and Creel (1961) \cite{Boomsliter} which states that a consonance
    is based on the length of the overall period of a stimulus.
    They show that consonant intervals, based on simple integer
    ratios of fundamental frequencies, have shorter overall periods
    than do dissonant intervals.

    Indeed, as we obtain for our model, the higher integers are
    $m$ and $n$, the higher number of states ($m + n - 1$) the
    interneuron has against the pattern of $\rho_{1,2}(t)$ peaks. In
    fact, the sequence of the states repeats periodically in time with the
    period $T_{state} = (2\pi/\Omega_1)m =
    (2\pi/\Omega_2)n$, which is the period of phases coincidences
    of $\cos \Omega_1 t$ and $\cos \Omega_2 t$, i.e. the overall period.
    But the interneuron gets into each state randomly. So, for the
    high number of states (dissonance) it is necessary much time in order to recognize
    some regularity inherent to the output spike train. Conversely,
    in the case of consonant input, the same amount of
    the spike train statistics details can be acquired in shorter
    periods of observation. Thus, the consonant inputs are
    in the priority against the dissonant ones, from the analysis
    viewpoint.

\section{Discussion and conclusions}
    In the work we try to follow a signal propagating
    throughout the neural-like system. The second layer of the
    system doesn't allow applying the framework of Markovian
    processes. Nevertheless, we propose the qualitative analysis
    yielding the main result of the work:
    the analytical expressions and the consistent
    algorithm applicable for an investigation of the ISI
    statistics and its transformations.

    The proposed algorithm is ready to be used for quick
    estimations of output distributions because of step-like
    shapes of the functions called "$\Phi(\dots)$" and narrow peaks
    of FPTPDs $\rho_1(t)$ and $\rho_2(t)$.

    On the other hand, the found procedure is clear enough to be
    implemented in the widely used programming environments.
    In such an implementation it provides a
    rather precise approximation (see the Fig. \ref{fig:results}) of
    the output ISI distribution (ISID) given by the direct
    computer simulation of Eq. (\ref{general model}).
    We should also emphasize that in order to
    obtain a smooth curve of the ISID, using the direct numerical
    simulation of the system (\ref{general model}), it is
    necessary to consume much more temporal, soft- and hard-ware
    resources than in the case of a program implementation
    of the proposed algorithm usage.

    In the simple case of the auditory system model we
    are able to discover existence of some accords (a combination of two
    sinusoidal signals), which evoke ISIDs blurring very fast with
    propagation from one neural layer to another (Fig. \ref{fig:dissonants}).
    And in our study these accords are the same as the dissonant
    ones in music, i.e. the dissonant accords are the ones, which
    are not able to "survive" in the noisy neural environment
    after a number of interneurons layers.

    We also show that from the perceptional point of view the
    dissonant accord's ISI statistics needs more time to be
    collected in comparison with the consonant accord's one.
    The latter one evokes a sharp ISID's shape, which is able to
    "survive" a number of proposed transformations, i.e. the same
    algorithm is applicable in order to understand what happens to
    the consonant accords on deeper layers of the neural system.

    As it is easy to see the output
    ISID contains peaks corresponding to quasi-periodical spike
    generation at frequencies, which are absent in the input
    signal. So, it is possible and intersting to investigate the
    Ghost Stochastic Resonance phenomenon \cite{Balenzuela,
    Chialvo} in details for this model. However, the current paper
    is focused on the theoretical approach to the whole ISID picture
    shaping. All sophisticated tuning of coupling coefficients,
    input frequencies, and noise intensities can be performed
    separately in a sake of resonances investigation, and this
    analysis can be also augmented by results revealed from
    the proposed analytical approach.

    The obtained results may be applied also in the context of such
    recent studies as, for example, the stimulus reconstruction
    from neural spike trains, where the information transmission
    under the noise influence is investigated \cite{Nikitin}. The
    other suitable context of these results application is the
    continuous investigation of the neuron's behavior under the
    influence of a constant bombardment of inhibitory and excitatory
    postsynaptic potentials somehow resembling a background
    noise that is typical for functioning conditions of, for
    example, the neocortical neurons \cite{Luccioli}.

\bibliography{bibliography}

\begin{thebibliography}{15}
\expandafter\ifx\csname natexlab\endcsname\relax\def\natexlab#1{#1}\fi
\expandafter\ifx\csname bibnamefont\endcsname\relax
  \def\bibnamefont#1{#1}\fi
\expandafter\ifx\csname bibfnamefont\endcsname\relax
  \def\bibfnamefont#1{#1}\fi
\expandafter\ifx\csname citenamefont\endcsname\relax
  \def\citenamefont#1{#1}\fi
\expandafter\ifx\csname url\endcsname\relax
  \def\url#1{\texttt{#1}}\fi
\expandafter\ifx\csname urlprefix\endcsname\relax\def\urlprefix{URL }\fi
\providecommand{\bibinfo}[2]{#2}
\providecommand{\eprint}[2][]{\url{#2}}

\bibitem[{\citenamefont{Gammaitoni et~al.}(1998)\citenamefont{Gammaitoni,
  Hanggi, Jung, and Marchesoni}}]{Gammaitoni}
\bibinfo{author}{\bibfnamefont{L.}~\bibnamefont{Gammaitoni}},
  \bibinfo{author}{\bibfnamefont{P.}~\bibnamefont{Hanggi}},
  \bibinfo{author}{\bibfnamefont{P.}~\bibnamefont{Jung}}, \bibnamefont{and}
  \bibinfo{author}{\bibfnamefont{F.}~\bibnamefont{Marchesoni}},
  \bibinfo{journal}{Rev. Mod. Phys.} \textbf{\bibinfo{volume}{70}},
  \bibinfo{pages}{228} (\bibinfo{year}{1998}).

\bibitem[{\citenamefont{Gerstner and Kistler}(2002)}]{Gerstner}
\bibinfo{author}{\bibfnamefont{W.}~\bibnamefont{Gerstner}} \bibnamefont{and}
  \bibinfo{author}{\bibfnamefont{W.}~\bibnamefont{Kistler}},
  \emph{\bibinfo{title}{Spiking Neuron Models. Single Neurons, Populations,
  Plasticity}} (\bibinfo{publisher}{Cambridge University Press},
  \bibinfo{year}{2002}).

\bibitem[{\citenamefont{Longtin et~al.}(2008)\citenamefont{Longtin, Middleton,
  Cieniak, and Maler}}]{Longtin}
\bibinfo{author}{\bibfnamefont{A.}~\bibnamefont{Longtin}},
  \bibinfo{author}{\bibfnamefont{J.~W.} \bibnamefont{Middleton}},
  \bibinfo{author}{\bibfnamefont{J.}~\bibnamefont{Cieniak}}, \bibnamefont{and}
  \bibinfo{author}{\bibfnamefont{L.}~\bibnamefont{Maler}},
  \bibinfo{journal}{Mathematical Biosciences} \textbf{\bibinfo{volume}{214}},
  \bibinfo{pages}{87} (\bibinfo{year}{2008}).

\bibitem[{\citenamefont{Fishman et~al.}(2001)\citenamefont{Fishman, Volkov,
  Noh, Garell, Bakken, Arezzo, Howard, and Steinschneider}}]{Fishman}
\bibinfo{author}{\bibfnamefont{Y.~I.} \bibnamefont{Fishman}},
  \bibinfo{author}{\bibfnamefont{I.~O.} \bibnamefont{Volkov}},
  \bibinfo{author}{\bibfnamefont{M.~D.} \bibnamefont{Noh}},
  \bibinfo{author}{\bibfnamefont{P.~C.} \bibnamefont{Garell}},
  \bibinfo{author}{\bibfnamefont{H.}~\bibnamefont{Bakken}},
  \bibinfo{author}{\bibfnamefont{J.~C.} \bibnamefont{Arezzo}},
  \bibinfo{author}{\bibfnamefont{M.~A.} \bibnamefont{Howard}},
  \bibnamefont{and}
  \bibinfo{author}{\bibfnamefont{M.}~\bibnamefont{Steinschneider}},
  \bibinfo{journal}{J Neurophysiol} \textbf{\bibinfo{volume}{86}},
  \bibinfo{pages}{2761} (\bibinfo{year}{2001}).

\bibitem[{\citenamefont{Helmholtz}(1954)}]{Helmholtz}
\bibinfo{author}{\bibfnamefont{H.~L.~F.} \bibnamefont{Helmholtz}},
  \emph{\bibinfo{title}{On the sensations of tone}}
  (\bibinfo{publisher}{Dover}, \bibinfo{address}{New York},
  \bibinfo{year}{1954}).

\bibitem[{\citenamefont{Burkitt}(2006)}]{Burkitt}
\bibinfo{author}{\bibfnamefont{A.~N.} \bibnamefont{Burkitt}},
  \bibinfo{journal}{Biol. Cybern.} \textbf{\bibinfo{volume}{95}},
  \bibinfo{pages}{97} (\bibinfo{year}{2006}).

\bibitem[{\citenamefont{Balenzuela}(2005)}]{Balenzuela}
\bibinfo{author}{\bibfnamefont{P.}~\bibnamefont{Balenzuela}},
  \bibinfo{journal}{Chaos} \textbf{\bibinfo{volume}{15}},
  \bibinfo{pages}{023903} (\bibinfo{year}{2005}).

\bibitem[{\citenamefont{Lopera et~al.}(2006)\citenamefont{Lopera, Buldu,
  Torrent, Chialvo, and Garcia-Ojalvo}}]{Lopera}
\bibinfo{author}{\bibfnamefont{A.}~\bibnamefont{Lopera}},
  \bibinfo{author}{\bibfnamefont{J.}~\bibnamefont{Buldu}},
  \bibinfo{author}{\bibfnamefont{M.}~\bibnamefont{Torrent}},
  \bibinfo{author}{\bibfnamefont{D.}~\bibnamefont{Chialvo}}, \bibnamefont{and}
  \bibinfo{author}{\bibfnamefont{J.}~\bibnamefont{Garcia-Ojalvo}},
  \bibinfo{journal}{Phys. Rev. E} \textbf{\bibinfo{volume}{73}},
  \bibinfo{pages}{021101} (\bibinfo{year}{2006}).

\bibitem[{\citenamefont{Chialvo et~al.}(2002)\citenamefont{Chialvo, Calvo,
  Gonzalez, Piro, and Savino}}]{Chialvo}
\bibinfo{author}{\bibfnamefont{D.}~\bibnamefont{Chialvo}},
  \bibinfo{author}{\bibfnamefont{O.}~\bibnamefont{Calvo}},
  \bibinfo{author}{\bibfnamefont{D.}~\bibnamefont{Gonzalez}},
  \bibinfo{author}{\bibfnamefont{O.}~\bibnamefont{Piro}}, \bibnamefont{and}
  \bibinfo{author}{\bibfnamefont{G.}~\bibnamefont{Savino}},
  \bibinfo{journal}{Phys. Rev. E} \textbf{\bibinfo{volume}{65}},
  \bibinfo{pages}{050902} (\bibinfo{year}{2002}).

\bibitem[{\citenamefont{Plomp and Levelt}(1965)}]{Plomp}
\bibinfo{author}{\bibfnamefont{R.}~\bibnamefont{Plomp}} \bibnamefont{and}
  \bibinfo{author}{\bibfnamefont{W.~J.~M.} \bibnamefont{Levelt}},
  \bibinfo{journal}{Journal of the Acoustical Society of America}
  \textbf{\bibinfo{volume}{38}}, \bibinfo{pages}{548} (\bibinfo{year}{1965}).

\bibitem[{\citenamefont{Cox}(1967)}]{Cox}
\bibinfo{author}{\bibfnamefont{D.}~\bibnamefont{Cox}},
  \emph{\bibinfo{title}{Renewal theory}} (\bibinfo{publisher}{Chapman and
  Hall}, \bibinfo{year}{1967}).

\bibitem[{\citenamefont{Gardiner}(1985)}]{Gardiner}
\bibinfo{author}{\bibfnamefont{C.~W.} \bibnamefont{Gardiner}},
  \emph{\bibinfo{title}{Handbook of stochastic methods}}
  (\bibinfo{publisher}{Springer}, \bibinfo{year}{1985}), \bibinfo{edition}{2nd}
  ed.

\bibitem[{\citenamefont{Boomsliter and Creel}(1961)}]{Boomsliter}
\bibinfo{author}{\bibfnamefont{P.}~\bibnamefont{Boomsliter}} \bibnamefont{and}
  \bibinfo{author}{\bibfnamefont{W.}~\bibnamefont{Creel}}, \bibinfo{journal}{J.
  Music Theory} \textbf{\bibinfo{volume}{5}}, \bibinfo{pages}{2}
  (\bibinfo{year}{1961}).

\bibitem[{\citenamefont{Nikitin et~al.}(2007)\citenamefont{Nikitin, Stocks, and
  Morse}}]{Nikitin}
\bibinfo{author}{\bibfnamefont{A.}~\bibnamefont{Nikitin}},
  \bibinfo{author}{\bibfnamefont{N.~G.} \bibnamefont{Stocks}},
  \bibnamefont{and} \bibinfo{author}{\bibfnamefont{R.~P.} \bibnamefont{Morse}},
  \bibinfo{journal}{Phys. Rev. E} \textbf{\bibinfo{volume}{75}},
  \bibinfo{pages}{021121} (\bibinfo{year}{2007}).

\bibitem[{\citenamefont{Luccioli et~al.}(2006)\citenamefont{Luccioli, Kreuz,
  and Torcini}}]{Luccioli}
\bibinfo{author}{\bibfnamefont{S.}~\bibnamefont{Luccioli}},
  \bibinfo{author}{\bibfnamefont{T.}~\bibnamefont{Kreuz}}, \bibnamefont{and}
  \bibinfo{author}{\bibfnamefont{A.}~\bibnamefont{Torcini}},
  \bibinfo{journal}{Phys. Rev. E} \textbf{\bibinfo{volume}{73}},
  \bibinfo{pages}{041902} (\bibinfo{year}{2006}).

\end{thebibliography}

\end{document}